\documentstyle[12pt,aasms4]{article}
\begin{document}

\title{The polarization and beaming effect for GRBs}

\author{K.S. Cheng}
\affil{Department of Physics, The University of Hong Kong,
Pokfulam Road, Hong Kong, China\\ E-mail: hrspksc@hkucc.hku.hk}

\and
\author{J.H. Fan}
\affil{ Center for Astrophysics, Guangzhou Normal University, Guangzhou
510400,
 China\\ E-mail: jhfan@guangztc.edu.cn}

\and
\author{Z.G. Dai}
\affil{Department of Astronomy, Nanjing University, Nanjing, China\\
 E-mail: daizigao@public1.ptt.js.cn}

\begin{abstract}

Both observations and theoretical models suggest that the emissions in
gamma-ray bursts (GRBs) and the
afterglows are beamed. We argue that the recent polarization measured
in the afterglows gives further evidence of beaming in GRBs.
In this approach, we adopted the polarization-magnitude
relation of BL Lacertae objects to 4 GRBs with available
polarization measurements and found that the data  of the
4 GRBs are consistent with the relation of BL Lacertae objects.
This result suggests that the emissions in  GRBs are beamed
and there are similarities between GRBs and BL Lacertae
objects (or generally blazars). We suggest that the weak polarization
of GRBs may result from low percentage of orderly magnetic field in
the shock front.

\end{abstract}

%\end{keywords}

\section{Introduction}

It has been widely argued that gamma-ray bursts (GRBs) and afterglows 
are beamed. There is much observational evidence 
that the emission regions of GRBs as well as
afterglows must be moving relativistically (for a review see Piran 1999).
The emitted radiation is strongly beamed, and one can observe only 
a region with an opening angle of $\gamma^{-1}$ (where $\gamma$ is 
the Lorentz factor of the emission regions) off the line of sight.
Emission outside this narrow cone cannot be seen. These considerations 
have lead to the analytical speculations and numerical studies 
on the existence of jets in GRBs (Mao \& Yi 1994) and afterglows
(Rhoads 1997, 1999; Sari, Piran \& Halpern 1999; 
M\'esz\'aros \& Rees 1999; Panaitescu \& M\'esz\'aros 1999; 
Wei \& Lu 1999; Moderski, Sikora \& Bulik 1999; Dermer, Chiang 
\& Mitman 1999; Huang et al. 1999). Furthermore, 
the similarity in some of the observed properties 
between GRBs and blazars led to the speculation that jets also
appear in GRBs (Dermer \& Chiang 1998). In addition, jets appear in the
context of many models for the GRB energy source (e.g., Ruffert et al.
1997; Paczy\'nski 1998; Vietri \& Stella 1998; MacFadyen, Woosley
\& Heger 1999; Khokhlov et al. 1999; Wheeler et al. 1999; 
Cheng \& Dai 1999; Aloy et al. 1999). 

Recently, observational evidence for beamed afterglows has been reported.
First, the redshifts of GRB 990123 and GRB 990510 combined with their
observed fluences imply their isotropic energy releases of 
$\sim 3.4\times 10^{54}$ and $3\times 10^{53}$ ergs respectively 
(Kulkarni et al. 1999a; Harrison et al. 1999), which are much larger
than the energy budgets of the stellar-mass source models. 
Such enormous energetic budget requires that GRB radiation must be 
highly collimated in some GRBs. Second, the observed 
marked breaks in the light curves of the optical afterglows from these
GRBs have been interpreted as an observational signature for jets
(Kulkarni et al. 1999; Harrison et al. 1999; Stanek et al. 1999)
because, as argued by Rhoads (1997, 1999), Sari et al. (1999), 
and M\'esz\'aros \& Rees (1999), the lateral expansion and edge effects 
of a relativistic jet may result in an obvious steepening of the light 
curve of radiation from the jet, although the other explanation also 
exists (Dai \& Lu 1999a, 1999b). Third, the observed radio flares of 
these bursts have been believed to provide an independent and 
excellent indication of a jetlike geometry in GRBs 
(Kulkarni et al. 1999b; Harrison et al. 1999). As argued by
Waxman, Kulkarni \& Frail 1998),
the radio afterglow from a spherical fireball must rise to peak
flux on a timescale of a few weeks; but owing to the lateral expansion,
the radio afterglow from the forward shock of a jet must fade down
about few days after the burst (Sari \& Piran 1999). Therefore, 
the relative faintness of the observed late-time radio emission 
implies the existence of a jet. Finally, more importantly,    
the polarization observed from the afterglows gives further evidence 
for beamed emission in GRBs and afterglows 
(see Gruzinov 1999; Sari 1999; Hjorth et al. 1999). 
If GRBs  are similar to BL Lacertae objects,
then we believe that these two kinds of objects must have similar
properties. In this paper, we will discuss the relation between
the polarization and the optical magnitude for 4 GRBs, which are
the only GRBs with available data.

\section{Model}

Here, we summarize the main result of our previous paper
(Fan et al. 1997, 1999, Papers I \& II).
 The observed flux,
 $S_{j}^{ob}$, of a relativistic jet is related to its intrinsic flux,
 $S_{j}^{in}$, by $S_{j}^{ob} = \delta^{p}S_{j}^{in}$, where $\delta$,
 the Doppler factor of the jet, is defined by
 $\delta = [\Gamma(1-\beta cos\theta)]^{-1}$, $\beta$ is the velocity
 in units of the speed of the light, $\Gamma = (1 - \beta^{2})^{-1/2}$
 is the Lorentz factor, and $\theta$ is the viewing angle.  The value
 of $p$ depends on the shape  of the emitted spectrum and the detailed
 physics of the jet (Lind \& Blandford 1985), $p = 3 + \alpha$ is for
 a moving sphere and $ p = 2 + \alpha$  is for the case of a continuous
 jet, where $\alpha$ is the spectral index.  We consider a two-component
 model, in which the totally observed flux, $S^{ob}$, is the sum of
 an unbeamed part $S_{unb}$ and a jet flux, $S_{j}^{ob}$.
 Assuming that the intrinsic flux of the jet, $S_{j}^{in}$, is some fixed
 fraction $f$
 of the unbeamed flux, $S_{unb}$, i.e., $S_{j}^{in} = fS_{unb}$, we have
 $ S^{ob} = (1 + f\delta^{p}) S_{unb}$.  If the flux is not totally
 polarized in the jet, and it is not unreasonable to assume that the jet
 flux consists of two parts, namely, the polarized and the unpolarized,
 with the two parts being proportion to each other, i.e.,
 $S_{j}^{in} = S_{jp} + S_{jup}$, $S_{jp} = \eta S_{jup}$, where $\eta$
 is a coefficient which determines the polarization of the emission
 in the jet, then the observed optical polarization can be expressed as
\begin{equation}
P^{ob} = {\frac{(1 + f) \delta_{o}^{p}}{1 + f\delta_{o}^{p}}} P^{in}
\end{equation}
where intrinsic polarization, $P^{in}$ is defined by
\begin{equation}
P^{in} = {\frac{f}{1 + f}}{\frac{\eta}{1 + \eta}}
\end{equation}
and $\delta_{o}$ is the optical Doppler factor.

From relation (1), we can obtain following relation
\begin{equation}
P^{ob} = k~\delta^{p}~10^{0.4m^{ob}}
\end{equation}
which follows
\begin{equation}
{\frac{P^{ob}_{1}}{P^{ob}_{2}}} =
{\frac{\delta_{1}^{3+\alpha}}{\delta_{2}^{3+\alpha}}}~
10^{0.4(m^{ob}_{1}-m^{ob}_{2})}
\end{equation}
 for the observations of any two epochs, where $m^{ob}$ is the observed
magnitude and $k$ is the
 proportional constant.
 As mentioned above, the $\delta$ can be estimated from
 $S^{ob} = \delta^{3+\alpha}S^{in} + S_{unb}$, therefore
 the ratio, $\zeta = ({\frac{\delta_{1}}{\delta_{2}}})^{3+\alpha}$
 can be expressed as
$$ \zeta =
{\frac{S^{ob}_{1} - S_{unb}}{S^{ob}_{2} - S_{unb}}} >
{\frac{S^{ob}_{1}}{S^{ob}_{2}}}$$
 if we assume that neither $S^{in}$ or $S_{unb}$ is  changed
 and $S^{ob}_{1}$ is greater than $S^{ob}_{2}$.
 In this sense, the ratio $\zeta$ can be written in the form
 of $\zeta = ({\frac{S^{ob}_{1}}{S^{ob}_{2}}})^{\lambda}$, where
 the parameter $\lambda$ can be expected to be near unity since
 $\delta^{3+\alpha}S^{in}$ is usually much greater than $S_{unb}$.
 So, the ratio of polarizations (4) yields
\begin{equation}
{\frac{P_{1}}{P_{2}}} = 10^{0.4(\lambda - 1) \Delta m}
\end{equation}
Based on a 37-BL Lacertae-object sample, $\lambda$ is found to be
$1.30\pm0.07$ with a chance of probability of $2.0\times 10^{-5}$ 
(Fan, Cheng, Zhang 1999, Paper II). In this sense,
relation (4) can be expressed  in the form

\begin{equation}
log P(\%) = (0.12 \pm 0.02) m + c1 
\end{equation}
where $c1$ is a constant. Since GRBs likely show beaming as in
blazar, then similar relation should be expected to exist for GRBs. 

\section{Application to GRBs}

 Recently, polarization was measured from the afterglow of gamma ray
bursts (GRBs):
$P_{opt} = 1.7 \pm 0.2\%$ for GRB 990510 (Covino et al. 1999; Wijers et
al.1999);
$P_{opt} \le 2.3\%$ for GRB 990123 (Hjorth et al. 1999);
$P_{Radio} \le 19\%$ for GRB 980329 (Taylor et al. 1998);
$P_{Radio} \le 8\%$ for GRB 980703 (Frail et al. 1998).
These measurements likely suggest that the gamma-ray bursts are beamed
(Gruzinov 1999; Hjorth et al. 1999; Rhoads 1999;
 Panaitescu \& Meszaros 1998; Sari 1999; Sari et al. 1999). 

 The optical afterglow has been observed from many GRBs and found to be
 variable (see 
 Bloom et al. 1998;
 Castrio-Tirado et al. 1999a,b; 
 Djorgorski et al. 1997;
 Galama et al. 1999; 
 Garcia et al. 1998;
 Groot et al. 1998;
 Halpern et al. 1999;
 Kulkarni et al. 1999;
 Sahu et al. 1997;
 van Paradijs et al. 1997 and references therein
). 
 Since most optical data are made in R band, here we adopted the 
 R magnitude, which is obtained quasi-simultaneously with the 
 measurements of the 
 polarization.  The relevant 
 data are listed in Table 1, where Col. 1 gives that name of the
 GRB, Col. 2 and 3 the polarization measurements and the corresponding
 reference; Col. 4 and 5 the photometric data and the corresponding
 reference. 

\begin{table*}
\caption{ Polarization and photometric data for 4 GRBs}
\begin{tabular}{lcccc}
\hline\noalign{\smallskip}
 $Name$ & $ P (\%)$ & Ref & $m_{R}$ &Ref   \\
(1)     & (2)              & (3)  & (4)       & (5) \\
\noalign{\smallskip} \hline
GRB 980329  & $\le$ 21$^{\dagger}$  & T98  & $\geq$23& T98\\
GRB 980703  & $\le$ 8$^{\dagger}$   & F98  & $\geq$22& B98\\
GRB 990123  & $\le 2.3$             & H99  & 20.0    & C99\\
GRB 990510  & $ 1.6\pm0.2$          & Co99,W99 & 19.1    & Co99\\
\hline
\end{tabular}\\
%\end{table*}
Notes to the Table\\
$^{\dagger}$ polarization obtained in radio band.\\
B98: Bloom et al. (1998);
C99: Castro-Tirado et al. (1999b);
Co99: Covino et al. (1999);\\
F98: Frail et al. (1998);
H99: Hjorth et al. (1999);\\
T98: Taylor et al. (1998);
W99: Wijers et al. (1999)
\end{table*}

The relevant data are shown in Fig. 1, in which the solid line
represent $log P(\%) = 0.12 m_{R} - 2.06$. It is clear that the
data of GRBs are consistent with the result (the slope ) obtained
for BL Lacertae objects (Paper II).

\begin{figure}
\epsfxsize=9cm
$$
\epsfbox{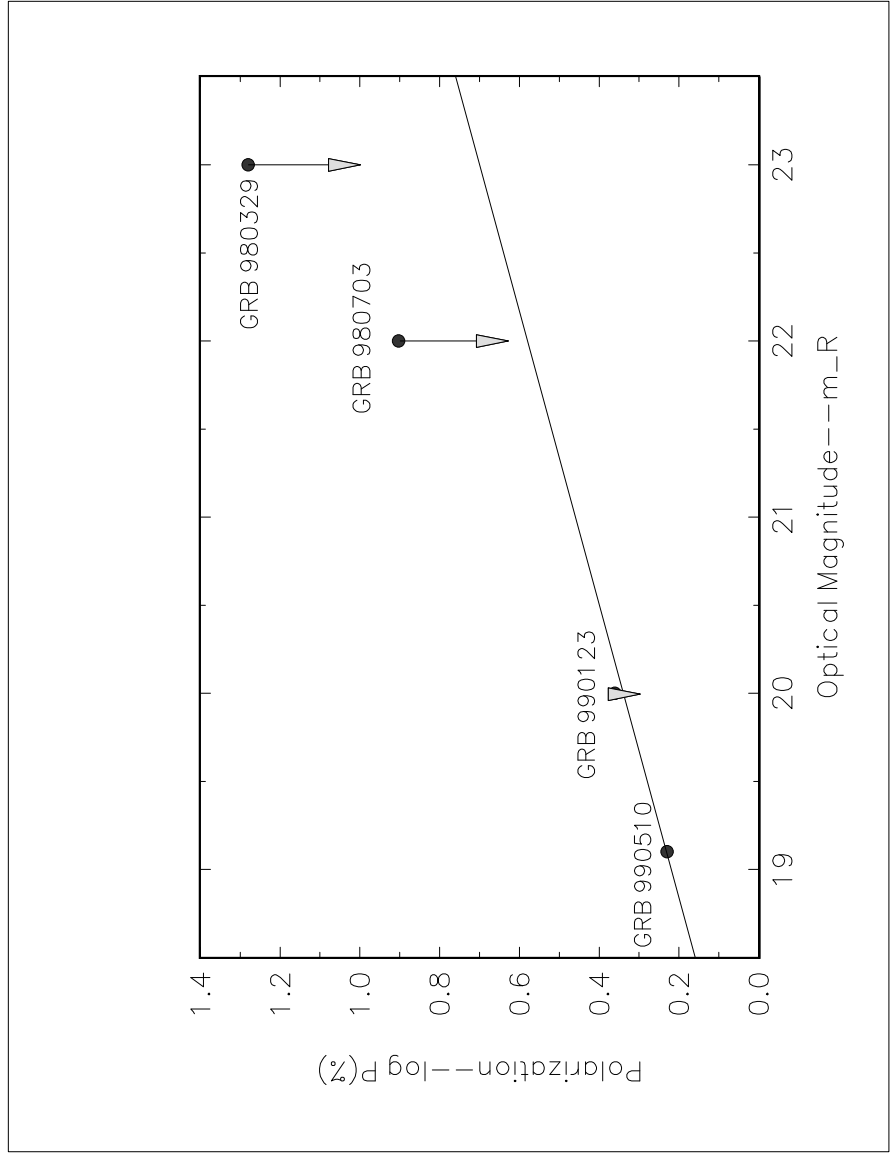}
$$
\caption {Plot of Polarization vs Magnitude--$m_R$ for 4 GRBs}
\label{fig:1}
\end{figure}

\section{Discussion}

 Relativistic fireball models can explain the main properties of the
 GRBs and their afterglows (Rees \& Meszaros 1992; Paczynski \&
 Rhoads 1993; Katz 1994; Meszaros \& Rees 1997; Vietri 1997;
 Waxman 1997; Sari et  al. 1998; Wijers \& Galama 1999).  
 Polarization can be observed if the fireball is beamed
 (Loeb \& Perna 1998; Waxman \& Gruzinov 1999;
 Hjorth et al. 1999). The polarization data of BL Lacertae objects
 indicates that the radio selected BL Lacertae objects (RBLs) have
 very high polarization $P~\sim~20\% - 40\%$ and X-ray selected
 BL Lacertae objects (XBLs) also have moderately high polarization,
 generally $P \leq 10\%$. In our previous papers (Fan et al. 1997, 1999,
 Paper I \& II), we found that the polarization of these BL Lacertae
 objects is associated with the Doppler factor and the 
 flux density, particularly, there is a correlation between the
 observed polarization and the optical magnitude. When we used
 the relation to the polarization of GRBs, they are consistent
 with the relation of BL Lacertae objects and this suggests that 
 (1) GRBs and BL Lacertae objects share similar properties and 
 (2) the observed polarization in the afterglow of GRBs indicate 
 that the emission is beamed.
 Sari (1999) and Ghisellini \& Lazzati (1999) found that, 
 the polarization value, assuming a highly beamed geometry, almost 
 linearly increases with time on a timescale of 1 day after GRB
 although two to three peaks appear in the whole evolution stage 
 of the polarization. Our taken data are about on 1 day after GRB
 (GRB 990123 and GRB 990510), 
 so the statistical result presented here is consistent with their 
 theoretical one near this time.   
 It has been shown that high polarization up
 to several tens percents is expected for GRBs if they are
 highly beamed (Loeb \& Perna 1998; Medvedev \& Loeb 1999;
 Waxman \& Gruzinov 1999; Gruzinov 1999). However, the
 observed polarization of GRB 990510 is only 1.6\%.  
 Apparently such low polarization value seems to indicate
 weak beaming factor. But there two factors to affect the
 polarization, namely beaming factor (boosting factor) and
 the orderly magnetic field in the shock front (intrinsic factor).
 Since the optical polarization from BL Lacertae objects is
 emitted from regions much closer to the central power house,
 typically $10^{15}$ cm, where the magnetic field is much orderly.
 On the other hand, the optical polarization of GRB afterglows is
 emitted from regions much farther away from the sources where
 magnetic field is produced by turbulence and hence should
 be more disoriented. In fact, it we assume that the minimum
 beaming factor of GRB 990510 is order of the unity and 
 $\Delta m \sim m^{ob}$ = 19.1, then $\delta \sim 100$, which
 is quite consistent with the fireball models.
 Since only one source, GRB 990510 shows definite polarization, 
 the constant, $c1 = -2.06$, in the relation
 $log P = 0.12 m_{R} - 2.06$ is obtained based on the data of 
 GRB 990510.
 If more polarization measurements are available, we would expect that
 the statistical coefficient of $m_{R}$ is the same, $0.12\pm0.02$
 but the constant $c1$ may be different.

\section*{Acknowledgements}
 We are grateful for the invaluable comments and suggestions from 
 Dr. A. Mitra. This work is partially supported by the RGC grant 
 of Hong Kong, the
 National Pan Deng Project of  China and the National Natural
 Scientific Foundation of China.

\newpage

Figure Caption\\

Fig. 1:  Plot of the polarization against the optical 
magnitude for GRBs with available data.

\end{document}